\documentclass[twocolumn,runningheads]{svjour2}
\smartqed  % flush right qed marks, e.g. at end of proof
\usepackage{graphicx}
\journalname{Astrophysics and Space Science}
\begin{document}

\title{The yields of r-process elements and chemical evolution of the Galaxy
}


\author{Zhe Chen  \and  Jiang Zhang \and YanPing Chen \and WenYuan Cui
\and   Bo Zhang }


\institute{Z. Chen \at
              Department of Physics, Hebei Normal University, Shijiazhuang 050016, P.R.China;
              Shijiazhuang Foreign Language School, 050000, P.R.China \\
%              Tel.: +123-45-678910\\
%              Fax: +123-45-678910\\
%              \email{fauthor@example.com}           %  \\
           \and
           J. Zhang \at
              Department of Physics, Hebei Normal University, Shijiazhuang 050016, P.R.China;
              Shijiazhuang University of Economics, Shijiazhuang 050031,
              P.R.China\\
           \and
           Y. P. Chen \and W. Y. Cui \and B. Zhang (corresponding
           author)\at
              Department of Physics, Hebei Normal University, Shijiazhuang 050016,
              P.R.China\\
              \email{ zhangbo@hebtu.edu.cn}
}

\date{Received: 22 Mar 2006 / Accepted: 09 Aug 2006}

\maketitle

\begin{abstract}
The supernova yields of r-process elements are obtained as a
function of the mass of their progenitor stars from the abundance
patterns of extremely metal-poor stars on the left-side
[Ba/Mg]-[Mg/H] boundary with a procedure proposed by Tsujimoto and
Shigeyama. The ejected masses of r-process elements associated
with stars of progenitor mass $M_{ms}\leq18M_{\odot}$ are
infertile sources and the SNe II with 20$M_{\odot}\leq M_{ms}\leq
40M_{\odot}$are the dominant source of r-process nucleosynthesis
in the Galaxy. The ratio of these stars 20$M_{\odot}\leq
M_{ms}\leq40M_{\odot}$ with compared to the all massive stars is
about $\sim$18\%. In this paper, we present a simple model that
describes a star's [r/Fe] in terms of the nucleosynthesis yields
of r-process elements and the number of SN II explosions. Combined
the r-process yields obtained by our procedure with the scatter
model of the Galactic halo, the observed abundance patterns
 of the metal-poor stars can be well reproduced.
%\PACS{First \and Second \and More}
\end{abstract}

\section{Introduction}
\label{intro} Elemental abundances in metal-poor Galactic halo
stars are providing evidence of the earliest Galactic
nucleosythesis history and clues about the identities of the first
stellar generations, the progenitors (or predecessors) of the halo
stars. The sample of identified field stars with [Fe/H]$\leq$-2.5
has increased by more than an order of magnitude in the last
decade. Because the elements in the atmospheres of these stars
have been produced in a small number of nucleosynthetic events,
abundance determinations can provide direct tests of model yields
from different nuclear processes.

The elements heavier than the iron peak are made through neutron
capture via two principal processes: the r-process and the
s-process (Burbidge et al., 1957). The r-process (for rapid
process) occurs when neutrons are added much more rapidly than the
$\beta$ decay times of the relevant nuclei. The site or sites of
the r-process are not known, although suggestions include the
$\nu$-driven wind of Type II SNe (e.g., Woosly and Hoffman, 1992;
Woosly et al., 1994) and the mergers of neutron stars (e.g.,
Lattimer and Schramm, 1974; Rosswog et al., 2000).

Particular attention to the Galactic evolution of elements
produced by neutron-capture nucleosynthesis was given by Mathews
et al., (1992), Pagel and Tautvaisiene (1997), and more recently
by Travaglio et al. (1999). These authors adopted the standard
approach to Galactic chemical evolution, assuming that stars form
from a chemically homogeneous medium at a continuous rate. A more
realistic model for the chemistry and dynamics of the gas is
needed in order to investigate the earliest phases of halo
evolution.

Recently, several chemical evolution models statistics on the
build-up of chemical elements in the early Galaxy (e.g., Tsujimoto
et al., 1999; Argast et al., 2000; Oey, 2000; Travaglio et al.,
2001; Fields et al., 2002); Tsujimoto et al., (1999) provided an
explanation for the spread of Eu observed in the oldest halo stars
in the context of a model of supernova-induced star formation.
Assuming that r-process nucleosynthesis sites would most likely be
identified with Type II supernovae (SNe II), Mathews et al. (1992)
suggested a mass range of $M_{ms}$=$7-8M_{\odot}$ for the site,
while Travaglio et al. (1999) supported a somewhat higher mass
range, $M_{ms}$=$8-10M_{\odot}$. Cescutti et al. (2005) concluded
that the Eu should originate as an r-process element in stars with
masses in the range $10-30M_{\odot}$.

Recent observations and the analyses imply that the abundance
pattern of an extremely metal-deficient star with [Fe/H]$\leq$-2.5
may retain information of a preceding single supernova (SN) event
or at most a few SNe (Mcwilliam et al., 1995; Ryan et al., 1996).
Tsujimoto and Shigeyama (1998) have also shown that the mass of
r-process elements ejected by each SN II as a function of
progenitor mass at the main sequence ($M_{ms}$) can be derived
from the observed [Ba/Mg]-[Mg/H] trend combined with the
[Mg/H]-$M_{ms}$ relation in theoretical SN models. But some of
extremely metal-deficient stars with [Fe/H]$\leq$-2.5 could be
formed out of gas enriched by several SNe (Fields et al., 2002).

In this paper, we assume that the stars on the left-side
[Ba/Mg]-[Mg/H] boundary are made from individual supernova events.
In Section 2 we discuss the r-process elements yields based on the
model proposed by Tsujimoto and Shigeyama (1998). In Section 3 the
r-process elements yields are used to explore the early stages of
inhomogeneous chemical evolution in the Galaxy. Conclusions are
given in Section 4.
\section{Production site and yield for r-process elements}
\label{sec:2}
%and \cite{Ref1}
Assuming that the stars on the left-side [Ba/Mg]-[Mg/H] boundary
retain the abundance pattern of a single supernova and using the
similar procedure presented by Tsujimoto and Shigeyama (1998), we
can calculate the r-process elements yields as a function of the
initial stellar mass from theoretical SN models (e.g., Woosley and
Weaver, 1995; Tsujimoto et al., 1995; Nomoto et al., 1997).

In Fig. 1a, which uses the data of Mcwilliam et al. (1995),
Mcwilliam (1998), Lai et al. (2004), Honda et al. (2004) and
Barklem et al. (2005), the [Ba/Mg] values for a sample of
metal-poor stars are plotted against [Mg/H]. We infer yields for
Ba element from the left-side boundary of observed abundances in
stars with -4$<$[Fe/H]$<$-2.5 by calculating the chi-squared fit
to the data. The fitted line is shown in Fig. 1a, which represents
the mass range of the r-process site. This gives a relation
between the metallicity [Mg/H] of stars and the mass Mms of SN II
progenitor as shown in Fig. 1b. Since the observed abundances in
CS22892-052 and other stars strongly enriched r-elements
([Eu/Fe]$\geq$1.0) may not reflect the composition of the ISM from
which it formed (Qian and Wasserburg, 2003, Barbuy et al. 2005),
we exclude these stars in our calculation.
\begin{figure}
\centering
  \includegraphics{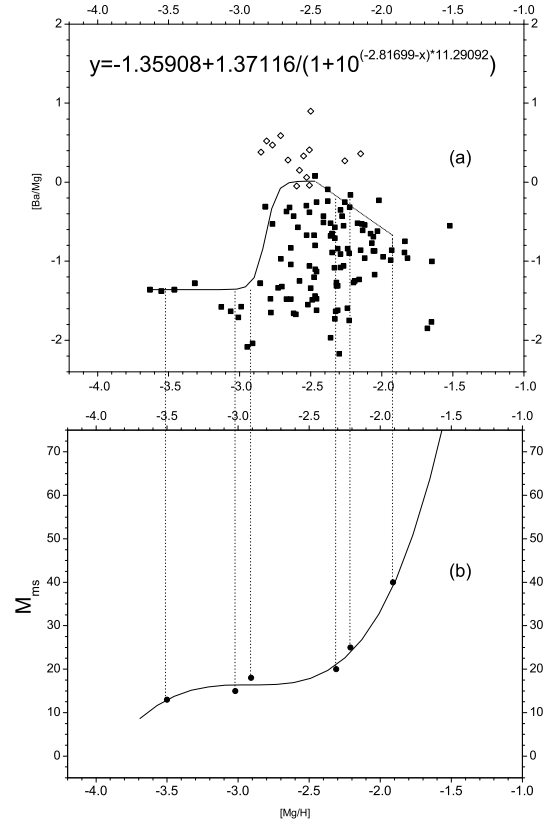}
\caption{(a) The correlation of [Ba/Mg] with [Mg/H] for
metal-stars; the thick line is the left-side boundary; the open
diamonds represent data of stars with [Eu/Fe]$\geq$1.0. (b)SN II
progenitor mass $M_{ms}$ is plotted as a function of [Mg/H] inside
the shell swept up by the SNR. Dashed lines in the lower and upper
panels illustrate how we assign $M_{ms}$ to each star along the
boundary.}
\label{fig:1}       % Give a unique label
\end{figure}

Using the theoretical nucleosynthesis mass of Mg for each Mms
(Tsujimoto et al., 1995), the ejected mass of element as a
function of $M_{ms}$ is derived. The yield of Ba obtained is shown
in Fig. 2. In addition, Eu is a tracer of the r-process, so we
investigate the enrichment of Eu as a representative of r-process
elements. Because the abundance pattern of heavy neutron capture
elements (Z$\geq$56) for each star is quite similar to that of the
r-process component in solar-system material (Senden et al., 1994,
Cowan et al., 1995), the Eu yield can be derived from
$M_{ej,Eu}=M_{ej,Ba}(\frac{Eu}{Ba})_\odot$and also shown in Fig.
2. According to our present models, SNe II with
$M_{ms}=12-20M_\odot$ do not produce significant amounts of
r-process elements. It is indicated that SNe with
$M_{ms}\geq18M_\odot$ are r-process main sites. The derived mass
of Ba synthesized in SNe II is $3\times10^{-6}M_\odot$ for
$M_{ms}=20M_\odot$, which is in good agreement with the values of
Tsujimoto and Shigeyama (2001) and Pastorello (2005). We remind
the reader that our model is based on the observed abundances of
the metal-poor stars, so the uncertainties of those observations
will be involved in the model calculations. We note from Fig. 1a
and 1b that for the massive stars (M$>25M_\odot$) the boundary is
not explicit, so the uncertainties of the r-process yields for
these stars are larger than those of lower mass stars.
\begin{figure}
\centering
\includegraphics[width=0.50\textwidth]{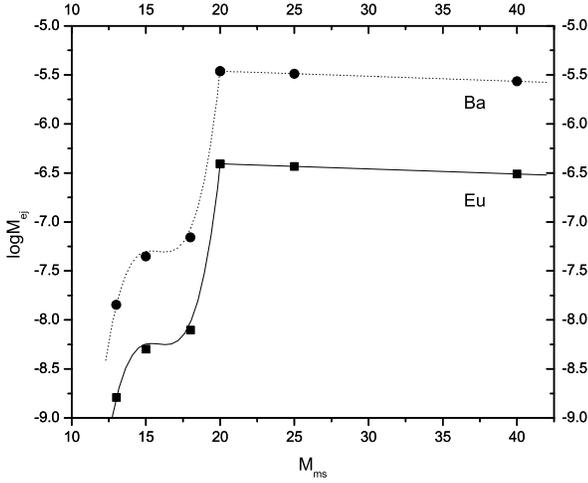}
\caption{Ejected masses of Ba and Eu from SNe II as a function of
the mass of their progenitor.}
 \label{fig:2}
\end{figure}
\section{The r-process scatter and the Galactic halo evolution}
%\subsection{Subsection title}
\label{sec:3} The very metal-poor stars, presently found in the
halo of the Galaxy, are believed to have formed at the ear liest
times, shortly after it became possible for the Universe to make
stars with sufficiently long main- sequence lifetimes to survive
for $\sim$14 Gyr (Senden et al., 2003). The chemical compositions
of these stars are thus expected to reflect a quite small number
of nucleosynthesis processes, possibly as small as one (e.g., the
stars on the left-side boundary in Fig. 1a), while the
compositions of metal-rich stars reflect the cumulative results of
the various processes that have been in operation during the
entire history of Galactic chemical evolution. Models for
r-process nucleosynthesis and for the Galactic chemical evolution
must account for the scatter of r/Fe between halo stars, as well
as for the smaller scatter between disk stars. We assume, as
Fields et al. (2002) did, that the observed Pop II r/Fe abundance
in each star reflect r-process contributions of a few SNe II. The
basic idea is that the r/Fe scatters arise from mixing among
different mass SNe II, which can produce different yields of
r-process elements.

To model the scatter of r/Fe, we create a set of halo stars and
deduce the history-the nuceosynthetic ancestry -of each. We assume
that a single supernova event mixes its metals and r-elements over
a region of mass M and analyze the mixing history of the
interstellar medium (ISM). For the first generation of SN, the
r-elements and iron yield of in the $i$th region is $m_{r1}$($i$)
and $m_{Fe1}$($i$) respectively. Since a given parcel of ISM gas
can be enriched to different degrees by the ancestors it had, the
new region (with mass M also) polluted by the successive
generation of SN comprises material originating from J old
regions. The accumulated r-elements and iron yields of $i$th new
region after the Nth generation of SN are
% as required. Don't forget to give each section and
%subsection a unique label (see Sect.~\ref{sec:1}).
%\paragraph{Paragraph headings} Use paragraph headings as needed.
\begin{equation}
m_{rN}(i)=m'_{r(N-1),i}+m_{rN}
\end{equation}
\begin{equation}
m_{FeN}(i)=m'_{Fe(N-1),i}+m_{FeN}
\end{equation}
where \ the \ reduced \ yields \ of \ the \ former \ (N-1) \
generations \ SNe \ contributed \ to \ the \ new \ region \ are
$m'_{r(N-1),i}=\sum_{j=1}^JP_{j,(N-1)}m_{r(N-1)}(k_j)$ and
$m'_{Fe(N-1),i}=\sum_{j=1}^JP_{j,(N-1)}m_{Fe(N-1)}(k_j)$. Here we
randomly take J group yields from the (N-1)th generation regions,
which are $m_{r(N-1)}(k_j)$ and $m_{Fe(N-1)}(k_j)$ (j=1, $\ldots$
,J) respectively. Random parameters $P_{j,(N-1)}$ (j=1, $\ldots$
,J) are the mass fractions (statistical weight) came from J old
regions to form the $i$th new region which will be polluted by Nth
generation SN. For $P_{j,(N-1)}$, $\sum_{j=1}^JP_j=1$.

To simplify notation, we will define the scale r/Fe ratio to be
\begin{equation}
R \equiv \frac{r/Fe}{(r/Fe)_\odot}
\end{equation}
which implies that [r/Fe]=logR When taking the yield of SN II with
$M_{ms}=20M_\odot$ and $m_{Fe}=0.07M_\odot$, we can obtain a value
of R$\approx$50, which is consistent with the value of Fields et
al. (2002). We assume that each of the halo stars we create
incorporates gas cloud which has been enriched by some number N of
supernova: this is the number of supernova ancestors for the star.
Let the total number of supernova events be N, we have that
\begin{equation}
R_{N,i}=\frac{m_{rN}(i)/A_r}{m_{FeN}(i)/A_{Fe}}/(r/Fe)_\odot
\end{equation}

At the very earliest time, when single supernova events really
contaminate a given star, the [r/Fe] scatter only populates the
extremes; at later time, mixing becomes efficient gradually, and
then the scatter decreases. So the [r/Fe] becomes a tracer of the
inhomogeneity of the halo. Several groups (Argast et al., 2000;
Tsujimoto et al., 1999; Oey, 2000) \ have \ used \ similar \
arguments to motivate detailed models that explain the observed
[r/Fe](specifically, [Eu/Fe]) scatter in terms of r-process
nucleosynthesis and an inhomogeneous chemical evolution of the
Galactic halo. In our model, the Eu yield is given in Fig. 2 and
iron yield is adopted as follows:\\
\emph{Case A}: $M_{Fe}=0.07M_\odot$\\
\emph{Case B}: $M_{Fe}$ is a function of progenitor mass
(Tsujimoto et al., 1995).

The metallicity determines the total number of supernova ancestors
via
\begin{equation}
[Fe/H]_{N,i}=logX_{Fe\ast}-log
X_{Fe\odot}=log\frac{m_{FeN}(i)}{M\times X_{Fe\odot}}
\end{equation}
where $X_{Fe\odot}$ and $X_{Fe}$ are the mass fraction of Fe in
solar system and the polluted region respectively. In the early
Galaxy, the gas composed of only hydrogen and helium with their
ratio $X_{H}$:$X_{He}$=0.75:0.25, so
\begin{equation}
M=M_{SW}/0.75
\end{equation}
where $M_{SW}$ is the mass of hydrogen swept up by an supernova
remnant (Shigeyama and Tsujimoto 1998). On one hand, when J is
large enough, equations (4) and (5) will give the form of
well-mixed chemical evolution of the Galactic halo which could not
explain the scatter of [Eu/Fe] observed for metal-poor stars, on
the other hand, when J=1, these equations will give the result
that all the Fe and r-elements ejected from N supernovae may
pollute only one supernova remnant and this is not what actually
occurs. In fact, the J must be very small, so we take J=2 in this
work.
\begin{figure}
\centering
  \includegraphics[width=0.55\textwidth]{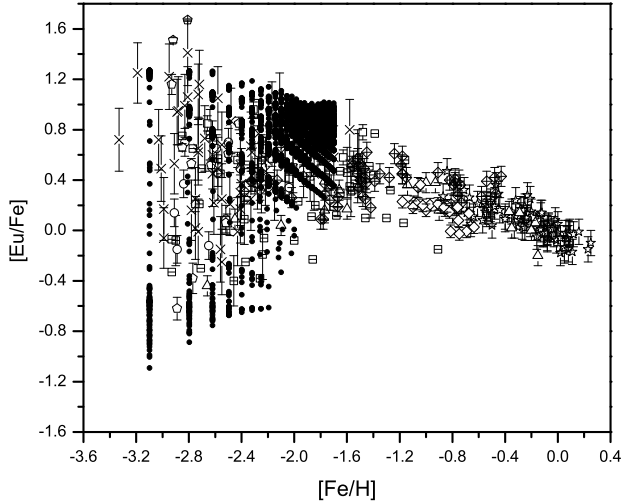}
\caption{Correlation of [Eu/Fe] with [Fe/H]. The symbols represent
the data taken from (open circles), Gilory et al. (1988) (+ center
squares) , Hartmann and Gehen (1988) (open hexagons) ,Zhao and
Magain (1990), Magain (1989) (open squares) , Da Siler et
al.(1990) (open down triangles) , Peterson et al. (1990) (pluses),
Edvardsson et al. (1993) (cross center squares) ,Gratton and
Sneden (1994) (open up triangles), Woolf et al. (1995) (open
stars) , Mcwilliam et al. (1995), Mcwilliam (1998) (open circles),
Shetrone (1996) (+ center diamonds), Jehin et al. (1999) (open
diamonds), Burris et al. (2000) (- center squares), Cowan et al.
(2002) (star), Honda et al. (2004) (open pentagons), Barklem et
al. (2005) (crosses). The dots represent the Case A result of our
model} \label{fig:3}
\end{figure}

It is of interest to consider making the mass of ancestor a random
variable. The initial mass distribution of supernova progenitor is
generated according to the formula of Eggleton et al. (1989),
\begin{equation}
M=\frac{0.19X}{(1-X)^{0.75}+0.032(1-X)^{0.25}}
\end{equation}
where X is a random number uniformly distributed between 0 and 1,
which leads to a mass function similar to that of Miller and Scalo
(1979). Results for a Monte Carlo simulation of a stellar
population appear in Fig. 3 and Fig. 4 for Case A and Case B
respectively. The degree of scatter increases with de creasing
metallicity because of counting statistics, so that the lowest
metallicity events record the nucleosyn thesis of a few events.
Using the abundance pattern of heavy neutron-capture elements
(Z$\geq$56) is quite similar to that of the r-process component in
solar-system material (Sneden et al., 1994; Cowan et al., 1995),
Figs. 5 and 6 show the scatters of [Ba/Fe], [Ce/Fe], [La/Fe],
[Nd/Fe], [Pr/Fe], [Sm/Fe] for Case A and B respectively. A
comparison with the observed data in Figs. 3-6 shows that the
model gives a good fit to the available data. We thus conclude
that the observed scatter in [r/Fe] can be understood by our
model.
\begin{figure}
\centering
  \includegraphics[width=0.55\textwidth]{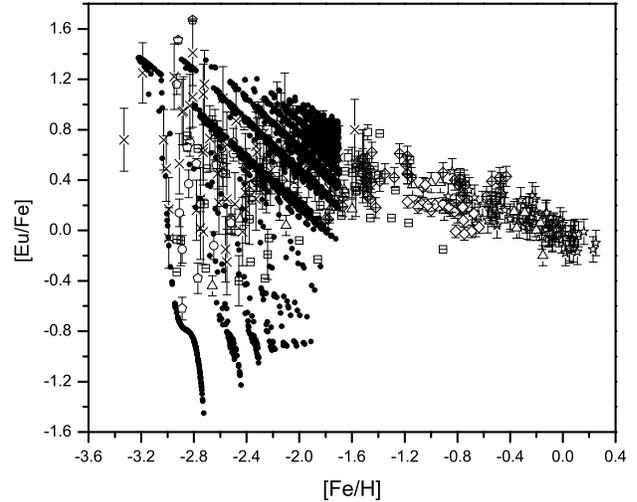}
\caption{Correlation of [Eu/Fe] with [Fe/H]. The symbols represent
the data taken from the same refs of Fig. 3. The dots represent
the Case B result of our model.} \label{fig:4}
\end{figure}
\begin{figure*}
\centering
  \includegraphics[width=0.75\textwidth]{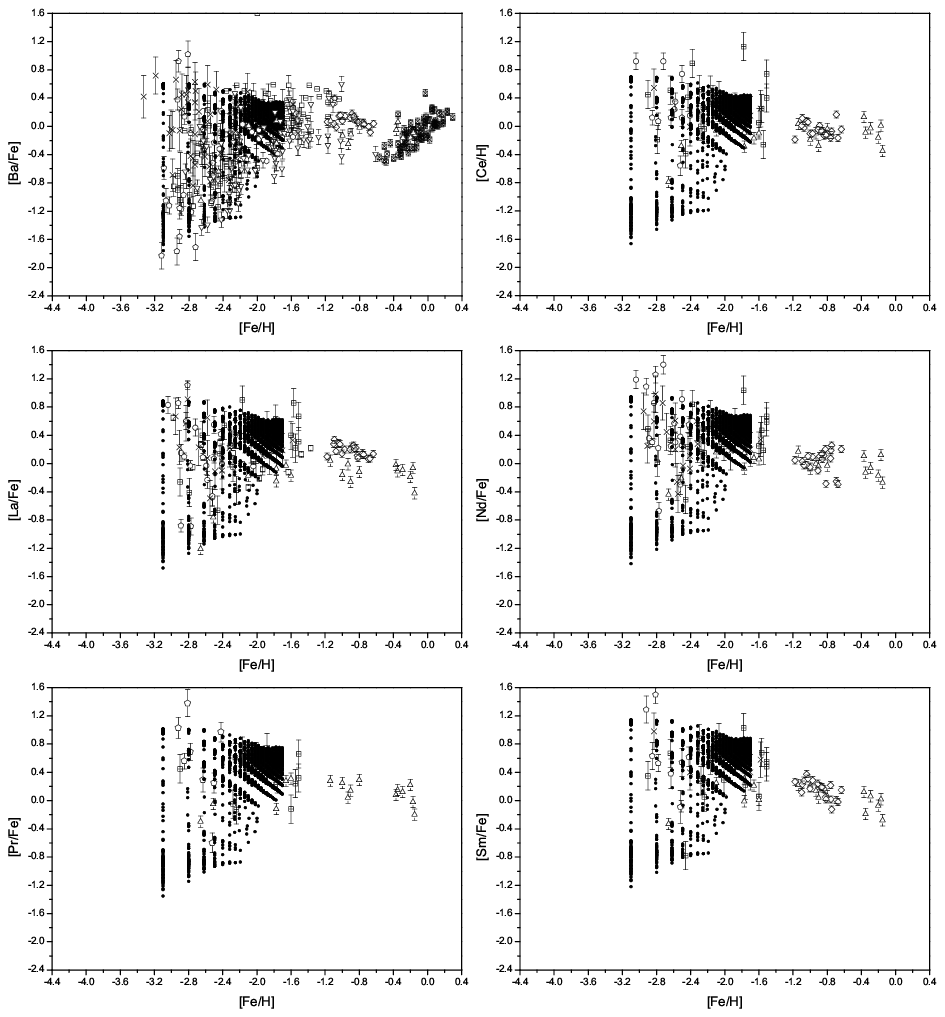}
\caption{Correlation of [X/Fe] with [Fe/H]. The symbols represent
the data taken from the same refs of Fig. 3. The dots represent
the Case A result of our model. Where X represents successionally
Ba, Ce, La, Nd, Pr, Sm.}
\label{fig:5}       % Give a unique label
\end{figure*}
\begin{figure*}
\centering
  \includegraphics[width=0.75\textwidth]{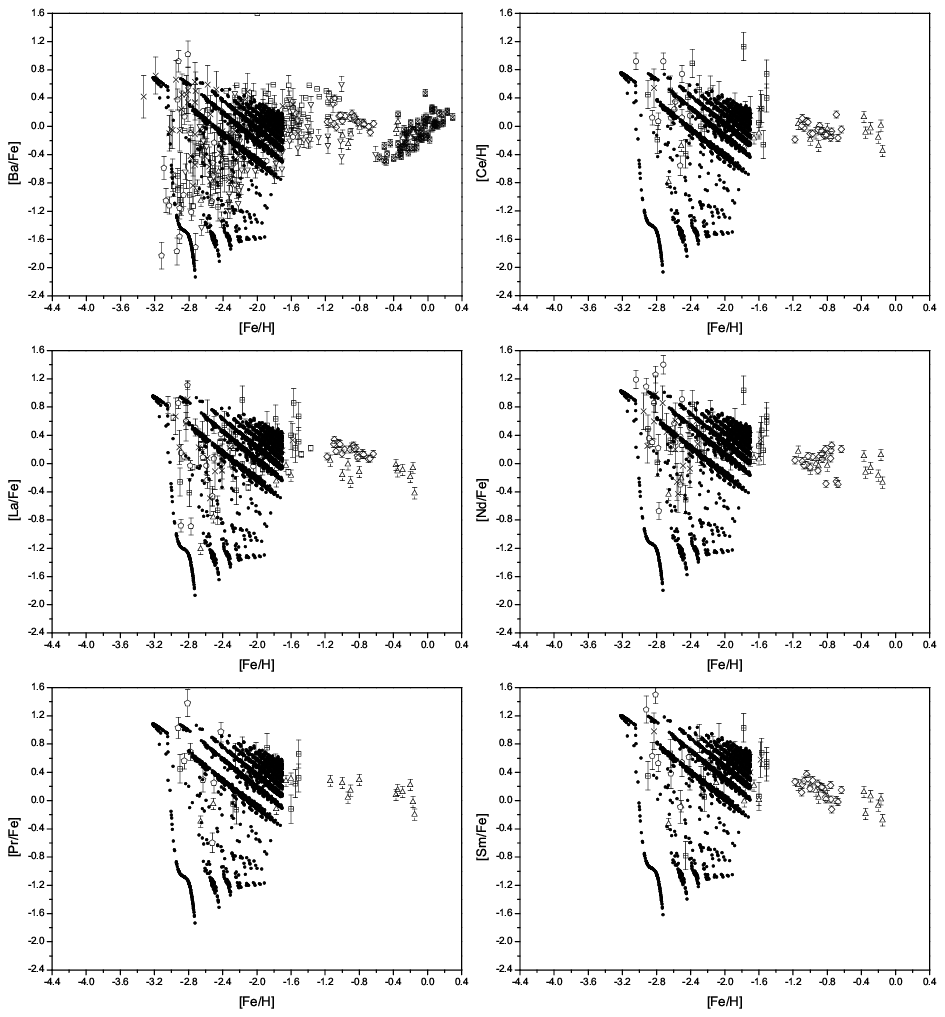}
\caption{Correlation of [X/Fe] with [Fe/H]. The symbols represent
the data taken from the same refs of Fig. 3. The dots represent
the Case B result of our model. Where X represents successionally
Ba, Ce, La, Nd, Pr, Sm.}
\label{fig:5}       % Give a unique label
\end{figure*}
\section{Conclusions}
\label{sec:4} Assuming that the stars on the left-side
[Ba/Mg]-[Mg/H] boundary retain the abundance pattern of a single
supernova, we have calculated the r-process elements yields as a
function of the initial stellar mass from theoretical SN models.
The scatter of [r/Fe] in halo strongly suggests that SNe II
associated with stars of progenitor mass $M_{ms}\leq18M_{\odot}$
are infertile sources for the production of r-process elements,
which is in agreement with the result of Tsujimoto et al. (2000).
We conclude that SNe II with $20M_{\odot}\leq M_{ms}\leq
40M_{\odot}$ are the dominant source of r-process nucleosynthesis
in the Galaxy. The effect on the SNe II which produces the high
r-process yield could be significant. Assuming a Salpeter IMF with
$\alpha$=-2.35, the ratio of these stars compared to the all
massive stars is about $\sim$18\%, which is higher than the value
of case A in Fields, Truran, and Cowan (2002). In fact, the mean
value of [r/Fe] obtained in our work is also higher than the value
of \={R} in Fields et al. (2002).

We have presented an approach to understand the scatter in heavy
r-process-to-iron ratio in metal-poor halo stars and used a
stochastic description of progenitor mass SNe II that contribution
to the heavy r-process and iron abundances in each halo star. The
random star-to-star variations in nucleosynthetic ancestry lead to
scatter in [r/Fe]. The models we present are all successful in
reproducing the scatter in the available data, which go down to
about [Fe/H]$\approx$-3. In conclusion, the ratios of [r/Fe] with
the metallicity are twofold. One is the abundance ratios for
[Fe/H]$\leq$-2.5 imprinted by the nucleosynthesis in a few
supernovae on the timescale $\sim10^{7}$ yr and the other for
[Fe/H]$\geq$-2 results from the mixing of the products from a
whole site of the nucleosynthesis, taking place on the timescale
longer than $10^{9}$ yr.
\begin{acknowledgements}
We are grateful to the referee for very valuable comments and
suggestions that improved this paper greatly. This work has been
supported by the National Natural Science Foundation of China
under grant No. 10373005.
\end{acknowledgements}

% BibTeX users please use
%\bibliographystyle{spmpsci}


\end{document}